\documentclass{article}
\usepackage{spconf,amsmath,graphicx}
\usepackage{amssymb}
\usepackage{url}
\usepackage{comment}
\usepackage{algpseudocode}
\usepackage{algorithm}
\usepackage{float}

\usepackage{xcolor}

\usepackage{makecell}

\title{End-to-end multi-talker audio-visual ASR using an active speaker attention module}
\name{Richard Rose and 
	Olivier Siohan}
\address{Google Inc., New York \\ \small \tt \{rickrose, siohan\}@google.com}

\begin{document}
\ninept
\maketitle
\begin{abstract}
This paper presents a new approach for end-to-end audio-visual multi-talker speech recognition. The approach, referred to here as the visual context attention model (VCAM), 
is important because it uses the available video information to assign decoded text 
to one of multiple visible faces.  
This essentially resolves the label ambiguity issue associated with most 
multi-talker modeling approaches which can decode multiple label strings
but cannot assign the label strings to the correct speakers.
This is implemented as a transformer-transducer based end-to-end model and evaluated
using a two speaker audio-visual overlapping speech dataset created from YouTube videos.
It is shown in the paper that the VCAM model improves performance with respect to previously
reported audio-only and audio-visual multi-talker ASR systems.
\end{abstract}

\section{Introduction}

This paper presents the multi-talker (M-T) visual context attention model (VCAM), an end-to-end (E2E) audio-visual (A/V) 
modeling approach for transcribing utterances in scenarios where
there is overlapping speech from multiple talkers.
The presence of overlapping speech in utterances arising from human-human interaction 
has been studied in several domains including meetings~\cite{Cetin2006} and call center tasks~\cite{Tripathi2020}.
In a study of interactions in an example call center domain, 
roughly 12\% of the word occurrences in client--operator interactions
were found to correspond to overlapping speech. 
Furthermore, this study showed that dramatically higher WERs for both machine and
human transcription were obtained in regions where talkers overlap.

In general, multi-talker models attempt to improve speech recognition from multiple 
overlapping speakers by decoding transcriptions from each speaker, as opposed to 
improving speech recognition from a target speaker in the presence of background speech.
The A/V approach presented here assumes that audio and video signals
are available for overlapping talkers. 
Visual features, in the form
of mouth tracks aligned with speech, are similar to those used in
A/V end-to-end models for single-talker ASR~\cite{Afouras-2018,petridis2018audiovisual,Makino2019}.
Face tracking and video processing are performed offline in this work using tools described
in~\cite{Makino2019}.

In audio-only decoding of multiple transcriptions from overlapping speech, there is a basic ambiguity associated with assigning transcriptions to speakers~\cite{Yu-2017,weng2015}.  However, it is well known that human listeners use visual information to disambiguate utterances from multiple talkers in multi-party human-human interactions.   This has motivated the VCAM approach presented here. The VCAM model incorporates the visual channel to assign each of the decoded transcriptions to the face associated with each of the overlapping speakers in the utterance. Assuming that a speaker's face serves as a proxy for speaker identity, this obviates the need to have a separate module for dealing with the ambiguity associated with assigning transcriptions to speakers.

The approach presented here builds on previous work in mask-based end-to-end A/V multi-talker
modeling~\cite{Tripathi2020,rose2021}.
The VCAM has the same mask-based structure as used in the audio-only multi-talker model in~\cite{Tripathi2020}
where a conventional recurrent neural
network transducer (RNN-T) architecture is
extended to include a masking model for separation of encoded
audio features.
It also includes multiple label encoders to encode transcripts
from overlapping speakers. 
This audio-only multi-talker model was extended in~\cite{rose2021} to
incorporate visual features from overlapping speakers 
with audio features.
It was shown to significantly improve speaker disambiguation in ASR on overlapping speech
relative to audio-only performance.

The input to the VCAM M-T model is an utterance containing overlapping speech along
with mouth-tracks associated with all on-screen faces. 
A separate decoding pass is made for each available mouth track.
For a given decoding pass at each audio frame, an attention weighted combination 
of encoded visual frames is input to a mask model with the encoded audio frame.  
Visual frames from a given mouth-track and audio frame are weighted according to their
similarity to the audio frame. 
It will be shown in Section~\ref{sec-system_description} that the attention maps 
produced by the model provide an indication when a given speaker is talking
even during speaker overlap intervals.

There has been a great deal of recent work on audio-only
end-to-end approaches to multi-talker ASR~\cite{kanda2020,Chang-2020,lu-2021}. 
All of these approaches involve an extension of the single label encoder end-to-end model
with a training procedure that aligns overlapping speech with
transcriptions from multiple speakers.
Further work has addressed the latency issues associated with multi-talker decoding~\cite{lu-2021}.
An E2E A/V M-T approach has recently been applied to addressing
the multi-speaker cocktail party effect~\cite{wu-2021}. 
There is also a large body of
work on speech separation where the goal is to recover
a target speech signal in the presence of background speech~\cite{hershey2015,Chen-2020,Yu-2017}.
This includes approaches that fuse audio andvisual features
for speech separation in videos~\cite{Ephrat2018,yu2020audiovisualicassp,Chao2016}. 
Explicit speech separation systems generally optimize criteria related to
signal-to-background distortion and overall signal fidelity.

A simulated $M\!\!=\!\!2$ speaker overlapping speech A/V corpus was created for
training A/V multi-talker models from a YouTube A/V
corpus that was originally created for training A/V ASR models~\cite{Liao2013,Makino2019}.
Each utterance in the overlapped
speech corpus was created by combining two
utterances with randomly selected overlap intervals ranging from
one to five seconds. This corpus
is described in more detail in Section~\ref{sec-study}.
The results of an experimental study is presented in Section~\ref{sec-results}.
WERs for end-to-end
multi-talker models are compared when evaluated on the simulated A/V overlapping
utterances described in Section~\ref{sec-study}.

\section{System Description}
\label{sec-system_description}
This section introduces the end-to-end visual context attention model (VCAM) approach
to multi-talker modeling. 
First, the audio-only and A/V M-T models presented in~\cite{Tripathi2020} and~\cite{rose2021} are reviewed.
Second, the VCAM A/V M-T approach is motivated and described. 
\vspace{-0.1in}

\subsection{Mask based A/V multi-talker model}
\label{subsec-AV-multitalker}

\begin{figure}[htbp]
	\centering
	\vspace{-0.1in}
	\hspace{-.15in}
	\includegraphics[width=9.0cm]{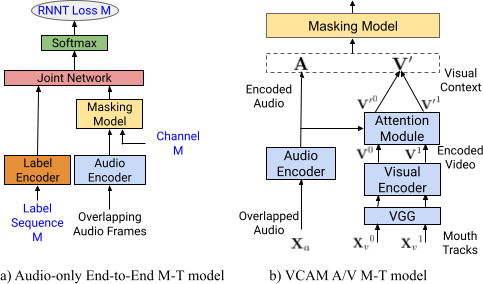}
	\vspace{-0.2in}
	\caption{a) Audio-only multi-talker (M-T) RNNT model. b) Visual context attention model (VCAM)
	which also includes label encoder and joint network.}
    \vspace{-0.05in}
	\label{fig-multitalker_review}
\end{figure}

A simplified block diagram of the audio-only multi-talker (M-T) model,
originally presented in~\cite{Tripathi2020}, is shown in
Figure~\ref{fig-multitalker_review}a.
It was shown in~\cite{Tripathi2020} that the single label encoder RNNT can be
extended to the multi-talker case by adding an LSTM masking
model as shown in the figure. It is assumed in the figure that
the audio input can contain up to M overlapping utterances. In
training, it is assumed that a separate reference label sequence
exists for each of the $M$ overlapping utterances from distinct speakers. Multi-talker
training is performed by separately aligning the overlapped audio
frames to each of the $M$ label sequences.
A unique channel sequence index, $m$, is appended to the encoded audio features
for each label sequence before inputting the encoded audio to the
masking model. This serves to disambiguate speech associated
with label sequence $m$ from competing speech.
Separate RNNT losses are computed for each of the $M$ label sequences, and the overall RNNT loss is the sum of channel
specific RNNT losses.
\vspace{-.05in}

The audio-only M-T model in Figure~\ref{fig-multitalker_review}a was
was extended to an A/V M-T model by replacing the audio encoder with an A/V encoder accompanied by
an attention network~\cite{rose2021}.
For each individual video frame, the attention network in~\cite{rose2021} produces a weighted combination
of the $M$ video frames corresponding to each of the $M$ label sequences.
For both the audio-only and A/V models,
separate RNNT losses are computed for each of the $M$ label sequences.
The overall RNNT loss is the sum of channel specific RNNT losses.
All parameters for all models in~\cite{Tripathi2020} and~\cite{rose2021} are trained using audio signals
containing simulated overlapping speech utterances.

\vspace{-0.1in}
\subsection{Visual context attention M-T model (VCAM)}
\label{subsec-VCAM}

The VCAM multi-talker model is depicted for the M=2 speaker case by the block diagram in Figure~\ref{fig-multitalker_review}b.
This model is part of the larger end-to-end framework and two pass training procedure described in Section~\ref{subsec-AV-multitalker}.
However, there are several important aspects of this model that distinguish it from
the M-T models discussed above.
First, 
unlike the approach in~\cite{rose2021} which computes attention weights frame-by-frame over mouth-tracks,
attention weighting is performed over all video frames separately for each mouth-track.
For each audio frame ${x_{a}}_{t}, t=1, \ldots, T$ and for each mouth-track $m$,
the attention module generates an an attention weighted visual context vector 
${v'}_{t}^{m}, t=1, \ldots, T$, from the encoded video features $v_{t}^{m}$.

The second important aspect of the VCAM model is that it performs 
late stage integration of audio and visual features as opposed to
simply concatenating audio and visual features at the input to and A/V encoder.
This is thought to allow the model to accommodate some degree of asynchrony between the
audio and video channels.
Third, the time-varying channel-dependent attention weighted visual context
vector in the VCAM is appended to the input of the masking model rather than a
one-hot channel select vector that is
appended to the input of the masking model for the M-T model in Figure~\ref{fig-multitalker_review}a.
It is shown below that the attention weighting over time
that is performed in the VCAM model has the effect of 
providing an approximation of a speaker identity signal to the masking model.
\begin{figure*}[h]
	\centering
	\vspace{-0.3in}
	\hspace{-0.7in}
	\includegraphics[width=17cm]{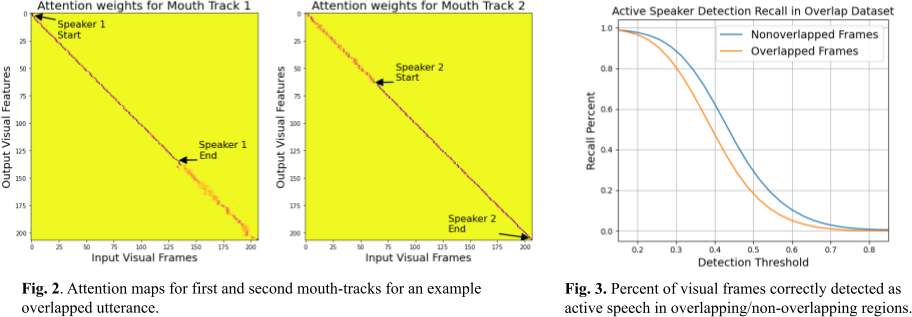} 
	\vspace{-0.1in}
    \vspace{-0.15in}
	\label{fig-vcam}
\end{figure*}

The VCAM M-T model is implemented as follows.
The audio features,
${\bf X}_{a} = \left \{{x_{a}}_{t} \right \}_{t=1}^{T}$,
for a T length utterance are $240$
dimensional vectors containing three stacked $80$ dimensional
mel-frequency filter-bank vectors extracted over 30 msec.\ frames.
The encoded audio vectors,
${\bf A} = \left \{a_{t} \right \}_{t=1}^{T}$,
are the $D_a = 1024$ dimensional output of a transformer-transducer (T-T)
with 6 layers and 8 attention heads~\cite{Zhang-2020}.

The input video frames,
${\bf X}_v\!\!=\!\!\left \{ {{x_v}}_{t}^{m} \right \}_{t=1, m=1}^{T, \;\;\;M}$,
for each of $M$ overlapping speakers
in a $T$ length utterance are $128\!\!\times\!\!128\!\!\times\!\!3$ thumbnail images.
Visual features are $512$ dimensional vectors computed from the input video
frames using a ``(2+1)D'' convolutional neural network~\cite{tran-2018}.
This convolutional network factorizes
3D convolution into
a 2D spatial convolution and a 1D temporal convolution.
It was found to achieve greater efficiencies than the 3D convolutional network
used in~\cite{rose2021} and~\cite{Makino2019} with no observed impact on WER.
Encoded visual features,
${V}^m\!\!=\!\!\left \{ {v}_{t}^{m} \right \}_{t=1}^{T}, m\!\!=\!\!1,2$,
are the $D_v\!\!=\!\!1024$ dimensional output of a T-T with 4 layers and 4 attention heads.

The attention network for the VCAM depicted in Figure~\ref{fig-multitalker_review}b computes
a $T\!\!\times\!T\!\times\!M$ dimensional tensor of attention weights 
based on the similarity between the encoded audio and visual feature vectors. 
The similarity between encoded audio at time $j$ and encoded video for mouth-track $m$
at time $i$ is computed as the inner product
\begin{equation}
S_{i,j}^{m} = {V_{i}^{m}}^{\mathsf{T}} A_{j}.
\label{eq-sim}
\end{equation}
The attention weights are the softmax similarity values:
\begin{equation}
w_{i,j}^m = Softmax(S_{i,j}^{m}).
\label{eq-softmax}
\end{equation}
The attention weighted visual context for the $m$th mouth-track,
${\bf V'}^m\!\!=\!\! \left \{{v'}_{t}^{m} \right \}_{t=1}^{T}$,
is obtained by attention weighting the encoded visual features across time:
\vspace{-0.1in}
\begin{equation}
\vspace{-0.05in}
{v'}_{j}^{m} = \sum_{i=1}^{T} w_{j,i}^{m} v_{i}^{m}.
\label{eq-attn}
\end{equation}
Finally, the encoded audio features and attention weighted encoded visual features are input
to a T-T based mask encoder with 6 layers and 6 attention heads and the $1024$ dimensional
mask output is input to the joint network of the E2E network.

\subsection{Detecting active speakers from VCAM attention maps}
\label{subsec-VCAM_ASD}
Figure~\ref{fig-vcam} displays two attention maps, one for
each of the two overlapping speakers' mouth-tracks in an example overlapped utterance.
The first overlapping waveform begins at frame $t\!\!=\!\!0$ and ends at frame $t\!\!=\!\!134$
with the attention weights associated with the mouth-track for that waveform shown in the plot on the left.
The second waveform
begins at frame $t\!\!=\!\!63$ and ends at $t\!\!=\!\!205$ with the attention weights for that
mouth-track shown in the plot on the right.
The figure illustrates the effect of the VCAM attention network.
Both plots display the $T \!\! \times \!\! T$ attention weights, $w_{i,j}^m, i,j\!\!=\!\!1,\ldots,T$,
from Equation~\ref{eq-softmax} which are applied to the input encoded visual frames $v_j^0$ and
$v_j^1$.
The $i$th horizontal line in each plot represents the weights applied to the 
$j\!\!=\!\!1,\ldots,T$ input visual frames at the $i$th audio frame.

There are two observations that can be made from these attention maps.
First, when a given talker is speaking, the attention weights for that 
speaker's mouth-track are extremely sparse, indicating that the correct 
mouth-track is being selected by the attention weighting.
Furthermore, this appears to be true even when there is overlapping speech (the interval between the
frames where speaker 2 starts talking and speaker 1 ends). 
Second, the magnitude of attention weights for a given mouth-track become
much smaller for those intervals where the associated talker is not speaking.

This implies that the attention module is effectively acting as a speaker detector, 
performing the role of implicit assignment of audio frames to the associated
on-screen face. 
This raises the question as to whether the VCAM model could be used for detecting active speakers in overlapped utterances.
The performance of active speaker detection (ASD) with this model is addressed in Section~\ref{sec-results}
for the overlapped speech dataset described in Section~\ref{sec-study}.

\vspace{-0.1in}
\section{Experimental Study}
\label{sec-study} 
\vspace{-0.1in}

This section describes the experimental study for
evaluating the performance of the
A/V multi-talker models presented in Section~\ref{sec-system_description}.
This study was performed using simulated overlapping speech data sets because
we are not aware of any publicly available audio-visual overlapping speech corpora.
We are aware of the publicly available simulated audio-only speech corpus
described in~\cite{Chen-2020} developed from LibriSpeech utterances~\cite{panayotov-2015}.
However, the experimental work in recent audio-visual speech separation work that
we are aware of, including work presented in~\cite{Ephrat2018,yu2020audiovisualicassp,Chao2016},
has been performed using simulated synthetic mixtures of single talker utterances. 
Audio-visual multi-talker ASR experiments described in~\cite{wu-2021} are based on
simulated 2-talker mixtures created from randomly selected single talker utterances
in the LRS2 corpus.
The A/V multi-talker experiments described in Section~\ref{sec-results} are performed
using the simulated A/V overlapping speech training and test sets described in Section~\ref{subsec_corpus}. 
\vspace{-.1in}
\subsection{Simulated audio-visual overlapping speech corpora}
\label{subsec_corpus}

A training set of simulated two-speaker overlapped utterances was created
from a corpus of single speaker A/V YouTube utterances containing aligned
face and mouth tracks.
The methodology behind the collection of the A/V YouTube corpus can
be found in~\cite{Liao2013,shillingford2018,Makino2019,rose2021}.
The overlapped audio waveform was created by taking
two of the above single speaker utterances, offsetting one in time with respect to the
other, and adding the two audio signals.
The video portion consists of two mouth tracks
where one of the mouth tracks has been offset to be aligned with the 
corresponding offset audio signal.
However, shifting the video frames creates a situation where there are 
no video frames associated with a given speaker in those regions where 
that speaker is not speaking. 
To enforce that the faces of the two speakers are available
throughout the entire utterance,
video frames for non-speech regions were 
filled with forward-backward repetitions of video frames from that speaker.
This provides video that is from the same speaker, but that is not
synchronized with the audio.  

There is a large amount of recent published work investigating
the important issue of generalizing
audio-only M-T approaches to scenarios with a larger number of speakers and more arbitrary
turn-taking~\cite{Raj-2021,kanda2020,Sklyar-2021, VonNeumann-2021}.
However, we also found it important to maintain accuracy for the multi-talker models on both
overlapping speech as well as single speaker utterances.
To do this, overlapping speech training data was combined with single speaker utterances.
Two single speaker scenarios were investigated.  The first scenario assumes that,
while a single talker is speaking, there is also an additional 
non-speaking on-screen face.  
To simulate this scenario for a given target utterance, a mouth-track from a 
different randomly chosen utterance is added, without the associated audio,
to the target utterance (``TwoFace'' scenario).
A second scenario assumes that there is a single on-screen speaker with no 
additional on-screen faces. 
This is simulated by using blank thumbnails for the second mouth-track 
in the A/V utterance (``OneFace'' scenario). 

The offset used in shifting the audio signals was chosen to
provide overlap intervals randomly selected with a uniform distribution 
between 1 and 5 seconds. Each overlapped speech utterance was stored with two reference transcriptions, two mouth
tracks, and overlap interval start and end times. 
The resulting training corpus contains 20k hours of training data.
Half of the training set consists of A/V overlapping speech utterances
and half consists of TwoFace single speaker utterances described above.

Overlapped and single speaker test sets were obtained from
human transcribed utterances with aligned mouth tracks taken
from YouTube videos. 
The process of forming overlapped utterances
is the same as described above for the training set. 
The test sets all contain 3601 utterances with the overlapped test utterances 
ranging in length from 2.7 to 14.7 seconds,
and the single speaker test utterances ranging in length from 2.5 to 8.0 seconds.  
\vspace{-0.1in}
\subsection{AI Principles}

The work presented in this paper abides by Google AI Principles~\cite{AIPrinciples}.
By improving the robustness of
speech recognition systems, we hope to increase the reach of ASR
technology to a larger population of users, and use it to develop 
assistive technology.
The data and models developed in this work
are restricted to a small group of researchers
working on this project and are handled in compliance with
the European Union General Data Protection Regulation~\cite{GDPR}.

\vspace{-.1in}
\section{Experimental Results}
\label{sec-results}
\vspace{-0.1in} Results are presented here for multi-talker
end-to-end models presented in Section~\ref{sec-system_description}
evaluated on the overlapping and single speaker test sets described in Section~\ref{sec-study}. 
First, a WER comparison is made between
the VCAM A/V M-T model presented in Section~\ref{subsec-VCAM}
and the audio-only and A/V M-T models described in Section~\ref{subsec-AV-multitalker}. 
Second, active speaker detection performance using the VCAM model is evaluated on the
overlapped test set.
Third, the impact of using transformer-transducer (T-T) versus LSTM based encoders in M-T models is evaluated. 
Finally, the performance of the M-T models on the single speaker scenario described in
Section~\ref{sec-study} is presented.

Table~\ref{tab:multitlaker-attn-comparison} provides a comparison between the WERs obtained for the
A/V VCAM M-T model from Section~\ref{subsec-VCAM}, the audio-only and A/V M-T models
described in Section~\ref{subsec-AV-multitalker}, and the A/V audio-only single channel model.
All \mbox{M-T} model WERs represent a dramatic reduction relative to the SingleChan A/V
model WER when evaluated on the Overlap test set.  
WERs shown in the table for the A/V M-T models represent greater than 18\%
reduction compared to the audio-only M-T model. 
The VCAM model provides a 
3\% reduction in WER relative to the previous A/V M-T model.
\vspace{-.1in}
\begin{table}[htbp]
\vspace{-0.1in}
	\centering
	\caption{Comparison of WERs for T-T based SingleChan and 
	MultiTalker models on Single and Overlap test sets.}
	\label{tab:multitlaker-attn-comparison}	
    \vspace{0.0in}
	\begin{tabular}{|c|c|c|}
	   \hline
	  \multicolumn{3}{|c|}{\bf WERs for T-T based SingleChan and MultiTalker Models} \\\hline
		\bf Model  & \bf Test Set & \bf WER \\ \hline\hline
 		 SingleChan A/V & Single &  12.2 \\ \hline
 		 SingleChan A/V & Overlap & 37.3 \\ \hline
 		 MultiTalker Audio & Overlap & 17.8 \\ \hline
		 MultiTalker A/V  & Overlap    &  14.5 \\ \hline
         MultiTalker VCAM A/V & Overlap &  14.1 \\ \hline 
    \end{tabular}
\vspace{-0.15in}
\end{table}

Section~\ref{subsec-VCAM_ASD} discusses the use of the VCAM attention module for detecting active speech in overlapping
and non-overlapping speech frames.
Active speaker detection (ASD) performance is measured here for the mouth-tracks in the overlapped speech test set 
described in Section~\ref{sec-study} using mean average precision (mAP).
For each mouth-track in an overlapped utterance, active speech is detected for a given visual frame when the maximum
attention weight for that frame, as depicted in Figure~\ref{fig-vcam}, exceeds a threshold. 
The mAP score is generated from a precision-recall curve by varying this threshold over a range
and averaging the precision score over a range of recall values. 
Figure 3 shows the recall, or percentage of visual frames correctly detected as active speech,
plotted separately for overlapping and non-overlapping frames. 
It is clear from the figure that recall for
overlapped frames is at most 15\% less than non-overlapped frames at a given detection threshold.
A mAP score of 95.8\% was obtained on this set where 27.7\% of the frames correspond to overlapping speech.
Of course, this task is far more constrained than other ASD tasks~\cite{Roth2019}.
However, it is important to note that this performance is obtained in the context of heavily overlapped speech.

Previous work on end-to-end multi-talker models was reported using LSTM based 
audio-only and A/V encoders~\cite{Tripathi2020,rose2021}.
Since all end-to-end models implemented in this work are based on T-T
encoders, the LSTM and T-T implementations of these models are compared here.
Table~\ref{tab:encoder-comparison} shows these WER comparisons for audio-only and A/V configurations of
M-T models. 
In both cases, the T-T implementation reduced WERs with respect to the
LSTM implementation by approximately 16 percent.
\vspace{-.1in}
\begin{table}[htbp]
    \vspace{-0.1in}
	\centering
    \caption{Comparison of WERs using LSTM and T-T encoders for audio-only 
    (Audio) and audio-visual (A/V) 
	multi-talker (MultiTalker) models on 2 talker overlapped
	speech (Overlap) test sets.}
	\label{tab:encoder-comparison}	
	\begin{tabular}{|c|c|c|c|}
	   \hline
	  \multicolumn{4}{|c|}{\bf WERs for Models with LSTM and T-T Encoders} \\\hline
		\bf Model  & \bf Encoder & \bf Test Set       & \bf WER  \\ \hline\hline
        MultiTalker Audio & LSTM & Overlap & 21.4 \\ \hline
        MultiTalker Audio & T-T & Overlap & 17.8 \\ \hline
        MultiTalker A/V & LSTM & Overlap & 17.4 \\ \hline
        MultiTalker A/V & T-T & Overlap & 14.5 \\ \hline
	\end{tabular}
\end{table}

While the results presented in Table~\ref{tab:multitlaker-attn-comparison} demonstrate that the VCAM
M-T model is capable of dramatic reductions in WER relative to the single channel end-to-end model on 
overlapping speech, it is still necessary to evaluate the M-T on single speaker 
non-overlapping utterances.
A comparison between WERs obtained for single channel and VCAM multi-talker models for the two
single speaker A/V scenarios discussed in Section~\ref{sec-study} is presented in 
Table~\ref{tab:singlespeakerwers}.
For the TwoFace single speaker scenario, the WER for the VCAM M-T model increases by 6.1\%
relative to the baseline single channel A/V model.
For the OneFace scenario the WER for the VCAM represents a 9.8\%
increase relative to the single channel model.
It is assumed the larger increase in WER observed for the OneFace scenario is at least
partly due to the fact that there were no OneFace single speaker utterances in the training set.
The goal of future work is to overcome the degradation associated with this and other unseen conditions by further augmenting the training set.

\vspace{-.15in}
\begin{table}[htbp]
	\centering
	\caption{WER comparison of MultiTalker and SingleChan models on single speaker utterances
	under two scenarios for simulating second on-screen face: 1) Simulated second on-screen face (TwoFace), and 2) No second on-screen face (OneFace).}
	\label{tab:singlespeakerwers}	
    \vspace{0.0in}
	\begin{tabular}{|c|c|c|}
	   \hline
	  \multicolumn{3}{|c|}{\bf Audio-visual RNNT Model Performance (WER\%)} \\\hline
		\bf A/V Model  & \bf OnScreenScenario  & \bf WER \\ \hline\hline
 		 SingleChan & N/A     &  12.2  \\ \hline
        MultiTalker VCAM & TwoFace    &  13.0 \\ \hline
        MultiTalker VCAM & OneFace &  13.4 \\ \hline 
	\end{tabular}
	\vspace{-0.15in}
\end{table}
%

%
\section{Summary and Conclusions}
\label{sec-conclusions}
\vspace{-0.1in}

This paper has presented the VCAM A/V end-to-end multi-talker ASR model.
The method avoids the label ambiguity problem in M-T models in that
it associates decoded transcriptions with a given on-screen
face.
The VCAM model was also shown to have the potentional of performing end-to-end
speaker diarization for overlapping speech along with multi-talker speech-to-text.
This is enabled by an attention module that is able to associate speaker dependent
mouth-tracks with one of multiple overlapping utterances.
Experimental results showed that the VCAM A/V M-T model obtained lower WER than previously
presented A/V M-T models on a 2 speaker simulated overlapping speech task.
Furthermore, the performance of the VCAM model for single speaker audio-visual utterances
represented only a small degradation with respect to single channel A/V models. 
Current work is directed towards evaluating VCAM performance on an A/V dataset
collected from meetings with multiple participants using a 360 degree camera and array microphone.

\vspace{-0.1in}
\section{Acknowledgments}
\vspace{-0.1in}
The authors would like to thank 
Takaki Makino and Hank Liao for their contributions to A/V speech corpus development,
Otavio Braga for work on efficient spatiotemporal convolutions for video analysis,
and Basi Garcia for help with tools in model evaluation.


\vfill\pagebreak
\ninept
\bibliographystyle{IEEEbib}
\bibliography{refs}
\end{document}